\documentclass[12pt,a4paper]{article}
\usepackage{cite}
\usepackage{amssymb}
\usepackage[final]{graphicx}

\newcommand{\dd}{\mathrm{d}}
\newcommand{\eps}{\epsilon}
\newcommand{\dilog}{\,{\mathrm{Li}}_2}
\newcommand{\Eep}{E_e{\!\!'}\,}

\newcommand{\GeV}{\ensuremath{\;\mathrm{GeV}}}
\newcommand{\Angle}{\measuredangle}

\begin{document}


\begin{titlepage}

\renewcommand{\thefootnote}{\fnsymbol{footnote}}
\setcounter{footnote}{0}

\begin{flushright}
  IKDA-01/22 \\
  SI-2001-7 \\
  hep-ph/0110120 \\
  October 2001
\end{flushright}

\vspace*{20mm}

\begin{center}
  {\Large\textbf{
    Deep Inelastic Scattering             \\[0.5ex]
    with Tagged Initial State Radiation:  \\[0.5ex]
    Complete $\mathcal{O}(\alpha)$ leptonic QED Corrections \\[0.5ex]
  }}
\end{center}


\vspace*{0.5cm}


\begin{center}
  {\large Harald Anlauf\,%
  \footnote{Supported by Bundesministerium f\"ur Bildung, Wissenschaft,
            Forschung und Technologie (BMBF), Germany.}%
  \footnote{Email: \texttt{anlauf@hep.tu-darmstadt.de}}
  } \\[1ex]
  \textit{%
	  Institut f{\"u}r Kernphysik,
	  Darmstadt University of Technology, \\
          64289 Darmstadt, Germany}
  \\[1ex]
  and
  \\[1ex]
  \textit{Fachbereich Physik, Siegen University,
          57068 Siegen, Germany}
\end{center}

\vspace*{0.5cm}

\begin{abstract}
  In this paper we extend the calculation of the QED corrections to deep
  inelastic lepton-proton scattering with a tagged photon, taking into
  account the full corrections on the lepton side.  Comparing to previous
  results that were obtained by considering only large logarithmic terms at
  leading and next-to-leading accuray, we find that the difference is in
  general quite small, however, it may be significant in the region of
  large $y$ and small $x$.
\end{abstract}

\begin{center}
  \textit{(Submitted to The European Physical Jounal C)}
\end{center}

\end{titlepage}


\setcounter{footnote}{0}



\section{Introduction}
\label{sec:Intro}

One of the major aims of the experiments at the HERA $ep$ collider is the
measurement of the structure functions of the proton, $F_2(x,Q^2)$ and
$F_L(x,Q^2)$, over a broad range of the kinematic variables.  Especially
the domain of small Bjorken $x < 10^{-4}$ and momentum transfer $Q^2$ of
the order of a few GeV$^2$ and below is of particular interest, as it
provides a challenge for attempts towards a complete, quantitative
understanding of the dynamics of quarks and gluons inside the nucleon.

Of these structure functions, the longitudinal one, $F_L$, is much more
difficult to access.  There exist several ways to separately extract $F_2$
and $F_L$ from the experimental data.  The most obvious, direct method
requires to run the collider at different (i.e., lower) center-of-mass
energies, which may not be desirable from the point of view of other parts
of the physics program.

Indirect methods usually require substantial input from theory and depend
more or less on modeling the hadronic final state, like extrapolations or
QCD fits (see e.g.,
\cite{DIS99:Arkadov,Adloff:2001qk,DIS01:Dubak,EPS01:ZEUS}), or the
measurement of the azimuthal angle distribution of final state hadrons, as
suggested by Gehrmann \cite{Gehrmann:2000vu}.  However, these methods can
be used at fixed collision energy.

Another direct method was suggested by Krasny et al.~\cite{KPS92} and
utilizes radiative events with an exclusive hard photon registered in the
forward photon detector (PD).  Such a device is actually part of the
luminosity monitoring system of the H1 and ZEUS experiments, and will
continue to exist after the HERA luminosity upgrade.  The idea of this
method is that emission of photons in a direction close to the incoming
electron corresponds to a reduction in the effective beam energy.  The
effective electron energy for each radiative event is determined from the
energy of the hard photon observed (tagged) in the PD.

Besides measuring $F_L$, radiative events extend the accessible kinematic
range to lower values of $Q^2$.  The potential of this method is supported
by preliminary results from the H1 collaboration of an analysis at low
$Q^2$ for $F_2$ \cite{Klein:1998mz,Issever:thesis,Issever:DIS01,H1:2001rad}
(for earlier analyses that did not take into account QED radiative
corrections see \cite{H1:rad,ZEUS96}).  The feasibility of the
corresponding determination of $F_L$ was studied in \cite{FGMZ96}.
However, with currently analyzed data sets it is not yet possible to
compete with $F_L$ from extrapolations or QCD fits \cite{Issever:thesis}.

A precise analysis of experimental data requires the inclusion of radiative
corrections.  The most important ones are the QED corrections on the lepton
side that have been discussed at the leading logarithmic level
\cite{BKR97,AAKM:ll} and taking into account next-to-leading logarithms
\cite{AAKM:JETP,AAKM:nlo,Anl99:Sigma}.  As the difference between leading
and next-to-leading logarithmic terms could be quite significant, reaching
of the order of 5 percent in some regions, it appears difficult to estimate
the remaining uncertainty due to non-logarithmic terms.  It is the purpose
of the present work to close this particular gap.

We may focus on the QED corrections on the lepton side since they form a
gauge invariant subset of the full corrections to the process under
consideration.  Furthermore, one expects that the QED corrections on the
hadronic side are significantly smaller, as the typical hadronic mass scale
of the order of the proton mass is much larger than the electron mass, so
there is no comparably large logarithm.  In fact, emission of photons off
quarks can essentially be absorbed into suitably defined parton
distribution functions, and the net effect turns out to be numerically
negligible \cite{Spiesberger:1995dm}.

The outline of the present paper is as follows.  After introducing our
notation in section 2, we devote section 3 to a discussion of the
individual contributions to the leptonic corrections with forward photon
tagging, and apply these corrections in section 4 to the case of radiative
deep inelastic scattering. In section 5 we present some numerical results
and our conclusions.  The appendices collect several formulae that are
useful for a numerical implementation of our calculation.


\section{Born cross section}

A a starting point, let us introduce our notation in the context of the
lowest order contribution to the radiative, semi-inclusive deep inelastic
scattering process
\begin{equation} \label{eq:tag-kin-born}
  e(p) + p(P) \to e(p') + \gamma(k) + X(P') \; ,
\end{equation}
where the photon $\gamma$ is assumed to be measured in the photon detector.

We choose invariant kinematic variables that use the measured, scattered
electron and take into account the energy loss due to photon emission
\cite{KPS92},
\begin{eqnarray}
\label{eq:kin-vars}
  \hat{Q}^2 & = & -(p-p'-k)^2 \; ,
  \nonumber \\
  \hat{x}   & = & \frac{\hat{Q}^2}{2P \cdot (p-p'-k)} \; ,
  \nonumber \\
  \hat{y}   & = & \frac{P \cdot (p-p'-k)}{P \cdot (p-k)} \; .
\end{eqnarray}

For HERA conditions, the polar angle $\vartheta_\gamma$ of the tagged
photon (as measured with respect to the incident electron beam) will be
very small, $\vartheta_\gamma \leq \vartheta_0$, with $\vartheta_0$ being
about 0.45~mrad in the case of the PD of H1.  We shall also assume below
that the scattering angle of the electron, $\theta$, is always much larger
than $\vartheta_0$.  Therefore, the energy fraction of the electron after
initial state radiation reads
\begin{equation} \label{def:z}
  z = \frac{2P \cdot (p - k)}{S} = \frac{E_e - E_\gamma}{E_e}
    = \frac{\hat{Q}^2}{\hat{x}\hat{y} S} \; ,
\end{equation}
where $E_e$ is the electron beam energy, $E_\gamma$ represents the energy
deposited in the forward PD, and $S = 2 P \cdot p$.

The relation between the shifted variables (\ref{eq:kin-vars}) and the
standard Bjorken variables of deep inelastic scattering reads:
\begin{equation}
	Q^2 = \frac{\hat{Q}^2}{z} \; , \qquad
	x = \frac{\hat{x}\hat{y}}{1 - z(1-\hat{y})} \; , \qquad
	y = 1 - z(1-\hat{y}) \; .
\end{equation}

The Born cross section, integrated over the solid angle of the photon
detector ($0 \leq \vartheta_\gamma \leq \vartheta_0$, $m/E_e \ll \vartheta_0
\ll 1$) takes a factorized form,
\begin{equation} \label{eq:Born}
  \frac{1}{\hat{y}} \,
  \frac{\dd^3\sigma_\mathrm{Born}}{\dd\hat{x}\,\dd\hat{y}\,\dd z} =
  \frac{\alpha}{2\pi} \, P(z,L_0) \, \tilde\Sigma(\hat{x},\hat{y},\hat{Q}^2) \; ,
\end{equation}
with
\begin{eqnarray} \label{eq:Sigma-tilde}
  \tilde\Sigma(\hat{x},\hat{y},\hat{Q}^2)
  & = & \frac{2\pi\alpha^2(-\hat{Q}^2)}{\hat{Q}^2\hat{x}\hat{y}^2}
  \Biggl[ 2(1-\hat{y}) - 2\hat{x}^2\hat{y}^2\frac{M^2}{\hat{Q}^2}
  \\ && \qquad \qquad {}
    + \left( 1+4\hat{x}^2\frac{M^2}{\hat{Q}^2}\right) \frac{\hat{y}^2}{1+R}
  \Biggr]
  F_2(\hat{x},\hat{Q}^2) \; ,
  \nonumber
\end{eqnarray}
and
\begin{eqnarray}
  P(z,L_0) & = & \frac{1+z^2}{1-z} L_0 - \frac{2z}{1-z} \; , \qquad
  L_0 = \ln\left(\frac{E_e^2\vartheta_0^2}{m^2}\right)  \; ,
  \nonumber \\
  \alpha(-\hat{Q}^2) & = & \frac{\alpha}{1-\Pi(-\hat{Q}^2)} \; ,
  \nonumber \\
  \label{eq:R}
  R & = & R(\hat{x},\hat{Q}^2) =
  \left( 1+4\hat{x}^2\frac{M^2}{\hat{Q}^2} \right)
  \frac{F_2(\hat{x},\hat{Q}^2)}{2\hat{x}F_1(\hat{x},\hat{Q}^2)} \, - 1
  \; .
\end{eqnarray}

The obvious advantage of the use of shifted variables (\ref{eq:kin-vars})
is their direct appearance in the argument of the structure functions in
the Born cross section.


\section{Leptonic corrections}

In the expression for the lowest order cross section, eq.~(\ref{eq:Born}),
we encountered the logarithm $L_0 =: \ln \zeta_0$ of the quantity
\begin{equation}
  \zeta_0 = \frac{E_e^2 \vartheta_0^2}{m^2} \; .
\end{equation}
Although $\vartheta_0 \ll 1$, we have $\zeta_0 \gg 1$ and $L_0 \approx 6.5
\gg 1$ for the conditions of the HERA PD's.  We shall therefore
consistently neglect contributions that are of order
$\mathcal{O}(\vartheta_0)$ or $\mathcal{O}(\zeta_0^{-1})$.

As explained in the introduction, the subset of leptonic QED corrections is
gauge invariant, and it also factorizes, thus allowing a discussion
isolated from the hadronic part.  Besides keeping things more transparent,
this also facilitates reusing the results in other calculations.

In the present section, we shall therefore consider the Compton subprocess
\begin{equation}
\label{eq:compton-process}
  e(p_1) + \gamma^*(-q) \to e(p_2) + \gamma(k) \; ,
\end{equation}
with the emission angle of the photon being integrated over the PD, while
taking into account the corrections from virtual and real QED corrections.
While performing this integration, we require that the remaining part of
the amplitude for the full process (i.e., $\gamma^*+p \to X$) depends only
weakly on the small transverse momentum of the forward photon.


\subsection{Compton tensor}

Let $M_\mu$ be the matrix element of the Compton scattering process
(\ref{eq:compton-process}), with the index $\mu$ describing the
polarization state of the virtual photon.  Adopting the notation of
\cite{KMF87}, we define the Compton tensor
\begin{equation}
\label{def:compton-tensor}
  K_{\mu\nu}
  = \frac{1}{(2e^2)^2} \sum_{\mathrm{spins}}
  M_{\mu}^{e\gamma^*\to e'\gamma}
 (M_{\nu}^{e\gamma^*\to e'\gamma})^* \; .
\end{equation}

Using current conservation, this tensor is conveniently decomposed as follows:
\begin{eqnarray}
\label{eq:compton-tensor}
  \nonumber
  K_{\mu\nu}
  &=& \frac{1}{2} \left( P_{\mu\nu} + P_{\nu\mu}^{*} \right)
  \; , \\
  P_{\mu\nu}
  &=& \tilde{g}_{\mu\nu}\left(B_g + \frac{\alpha}{2\pi}T_g\right)
  + \sum_{i,j=1,2} \tilde{p}_{i\mu}\tilde{p}_{j\nu}
    \left(B_{ij} + \frac{\alpha}{2\pi}T_{ij}\right)
  \; ,
  \\ \nonumber
  \tilde{g}_{\mu\nu} &=& g_{\mu\nu} - \frac{q_{\mu}q_{\nu}}{q^2}
  \; , \quad
  \tilde{p}_{i\mu} = p_{i\mu} - q_{\mu}\frac{p_i \cdot q}{q^2}
  \; , \quad
  i=1,2 \; .
\end{eqnarray}

The expressions for the quantities $B_{ij}$ corresponding to the Born
approximation are:
\begin{eqnarray}
\label{eq:compton-born}
  B_g
  &=& \frac{1}{\hat{s}\hat{t}}
      \left[(\hat{s}+\hat{u})^2 + (\hat{t}+\hat{u})^2\right]
      - 2m^2q^2\left(\frac{1}{\hat{s}^2} + \frac{1}{\hat{t}^2}\right)
  \; , \nonumber \\
  B_{11}
  &=& \frac{4q^2}{\hat{s}\hat{t}} - \frac{8m^2}{\hat{s}^2}
  \; , \quad
  B_{22}
  = \frac{4q^2}{\hat{s}\hat{t}} - \frac{8m^2}{\hat{t}^2}
  \; , \quad
  B_{12} = B_{21} = 0 \; ,
  \\
  \nonumber
  \hat{s} &=& 2p_2 \cdot k \; , \quad
  \hat{t} = - 2p_1 \cdot k \; ,
  \\
  \nonumber
  \hat{u} &=& (p_1 - p_2)^2 \; , \quad
  \hat{s}+\hat{t}+\hat{u} = q^2
  \; .
\end{eqnarray}

Note that for almost collinear emission, $k \simeq (1-z) p_1$, we may
neglect the transverse momentum of the emitted photon in the tensor
decomposition (\ref{eq:compton-tensor}) and use momentum conservation to
set $\tilde{p}_2 = z \tilde{p}_1$.

The kinematic variables of the Compton subprocess are related to those of
the radiative DIS process via:
\[
  \hat{u} = - \frac{\hat{Q}^2}{z}
  \; , \quad
  q^2 = (p_1-k-p_2)^2 \simeq - \hat{Q}^2
  \; , \quad
  \hat{s} \simeq \frac{1-z}{z} \, \hat{Q}^2
  \; .
\]

The quantities $T_g$, $T_{ij}$ in (\ref{eq:compton-tensor}) denote the
radiative corrections to the Compton tensor.


\subsection{Virtual and soft corrections}

The virtual corrections to the Compton tensor, as calculated in
\cite{KMF87}, are conveniently decomposed into a piece containing the
universal infrared singular contributions, which are proportional to the
Born contributions, and an infrared finite remainder:
\begin{equation}
  T_g = \rho B_g + T_g'
  \; , \quad
  T_{ij} = \rho B_{ij} + T_{ij}'
  \; , \quad
  i,j=1,2
  \; ,
\end{equation}
where
\begin{equation}
  \rho = 4\ln\frac{\lambda}{m}(L_Q-1) - L^2_Q + 3L_Q + 3\ln z
  +\frac{\pi^2}{3} -\frac{9}{2}
  \; , \quad
  L_Q = \ln\frac{Q^2}{m^2}
  \; .
\end{equation}
The parameter $\lambda$ in the above expression is a fictitious photon mass
regulating the IR divergency.

Performing the integration over photon angles, we obtain
\begin{eqnarray}
  \label{eq:virt-int}
  \frac{E_e^2}{\pi}
  \int \dd\Omega_k \;
  B_{\mu\nu}
  & = &
  \left( - Q_l^2 \tilde{g}_{\mu\nu}
         + 4z \tilde{p}_{1\mu} \tilde{p}_{1\nu} \right) \times
  \\
  && {} \frac{1}{1-z}
  \left[
    \left( 1 + \frac{\alpha}{2\pi} \rho \right) P(z,L_0)
	     - \frac{\alpha}{2\pi} T
  \right]
  + \mathcal{O}\left(\vartheta_0^2,\zeta_0^{-1}\right)
  \; ,
  \nonumber
\end{eqnarray}
where
\begin{eqnarray*}
  T &=&
  (A\ln z + B) P(z,L_0)
  + C L_0 + D \; ,
  \\
  A &=& 2 L_Q - L_0 - 2\ln(1-z) \; ,
  \\
  B &=& \ln^2 z - 2 \dilog (1-z) - \frac{1}{2} \; ,
  \\
  C &=&
  - \frac{2z}{1-z}\ln z
  - z \; ,
  \\
  D &=&
  - \frac{1-6z+4z^2}{1-z} \left( \dilog(1-z) + \ln z \ln(1-z) \right)
  \\
  &-& 2z \ln^2(1-z) + \frac{8z}{1-z} \ln z
  - \frac{4 \pi^2}{3} z + 1 \; ,
  \\
  \dilog(x) & = & -\int\limits_{0}^{x}\frac{\dd y}{y}\ln(1-y) \; .
\end{eqnarray*}
The single and double logarithmic terms in $L_0$ and $L_Q$ of the above
expression agree with ref.~\cite{AAKM:nlo}.  Details of the calculation
will be given elsewhere \cite{Anlauf:2002}.

The dependence of the virtual corrections on the unphysical parameter
$\lambda$ is canceled by the contribution from emission of an additional
soft photon, as usual.  Requiring that the energy fraction of the second
(soft) photon in units of the energy of the incoming electron does not
exceed $\eps$, with $\eps \ll 1$, and adding the contribution from soft
photon emission to the virtual correction then amounts to the replacement
of the quantity $\rho$ in (\ref{eq:virt-int}) by $\tilde\rho$, see
\cite{KMF87}:
\begin{eqnarray}
\label{eq:rho-v+s}
  \tilde{\rho} = 2(L_Q - 1)\ln\frac{\eps^2}{Y}
  + 3L_Q + 3\ln z - \ln^2Y - \frac{\pi^2}{3} - \frac{9}{2}
  + 2\dilog\left(\frac{1+c}{2}\right) \; ,
\end{eqnarray}
with
\begin{equation} \label{eq:Y-c}
  Y = \frac{\Eep}{E_e}
  \quad \mbox{and} \quad
  c = \cos\theta = \cos\Angle(\vec{p},\vec{p}\,{}')
\end{equation}
being the relative energy of the scattered electron and the cosine of the
scattering angle in the lab system, respectively.


\subsection{Double hard bremsstrahlung}

For the case of double photon emission, we define the `double Compton tensor'
analogously to (\ref{def:compton-tensor}) as
\begin{equation}
\label{def:double-compton-tensor}
  K_{\mu\nu}^{\gamma\gamma}
  = \frac{1}{2e^6} \sum_{\mathrm{spins}}
  M_{\mu}^{e\gamma^*\to e'\gamma\gamma}
 (M_{\nu}^{e\gamma^*\to e'\gamma\gamma})^* \; ,
\end{equation}
where now $M_{\mu}^{e\gamma^*\to e'\gamma\gamma}$ is the matrix element of the
double Compton process process
\begin{equation}
\label{eq:double-compton-process}
  e(p_1) + \gamma^*(-q) \to e(p_2) + \gamma(k_1) + \gamma(k_2) \; ,
\end{equation}
with the index $\mu$ describing the polarization of the virtual photon.

For the kinematic invariants of this subprocess we shall use the notation:
\begin{eqnarray*}
  z_i  & = & 2 p_1 \cdot k_i
  \; , \quad
  z_i' = 2 p_2 \cdot k_i
  \; , \\
  \sigma & = & 2 k_1 \cdot k_2 = (k_1+k_2)^2 \; , \\
  \Delta  & = & - [(p_1-k_1-k_2)^2 - m^2] = z_1 + z_2 - \sigma \; , \\
  \Delta' & = & [ (p_2+k_1+k_2)^2 - m^2 ] = z_1' + z_2' + \sigma \; , \\
  Q_l^2 & = & - (p_1-p_2)^2 = 2 (p_1 \cdot p_2 - m^2) \; , \\
  Q_h^2 & = & - q^2 = Q_l^2 + z_1 + z_2 - z_1' - z_2' - \sigma \; .
\end{eqnarray*}
%


\subsubsection{Double collinear emission}

When both photons are emitted almost collinearly to the incoming electron,
($k_1 \simeq x_1 p_1$, $k_2 \simeq x_2 p_1$), we may neglect the transverse
momenta of the photons with respect to the incoming electron, and the
double Compton tensor takes a simple form:
\begin{eqnarray}
  K_{\mu\nu}^\mathrm{2-coll} & = &
  4 \left[
  - \tilde{g}_{\mu\nu} Q_l^2
  + 4 z \left( \tilde{p}_{1\mu} \tilde{p}_{1\nu} \right)
  \right] \times
  \nonumber \\ &&
  \Biggl[
    \frac{1+z^2}{x_1 x_2} \frac{1}{z_1 z_2}
  - \frac{z}{\Delta^2} \left( \frac{z_1}{z_2} + \frac{z_2}{z_1} \right)
  + \frac{1}{x_1 x_2}
    \left( \frac{r_1^3+z r_2}{z_1 \Delta}
         + \frac{r_2^3+z r_1}{z_2 \Delta} \right)
  \nonumber \\ &&
  \; {}
  - 2 \frac{m^2}{\Delta}
    \left( \frac{r_1^2+z^2}{x_2 z_1^2}
         + \frac{r_2^2+z^2}{x_1 z_2^2}
         + \frac{(1-z)(r_1 r_2 + z)}{x_1 x_2 z_1 z_2}
    \right)
  \\ &&
  \; {}
  - 4 z\, \frac{m^2}{\Delta^2} \left( \frac{1}{z_1} + \frac{1}{z_2} \right)
  + 4 z\, \frac{m^4}{\Delta^2} \left( \frac{1}{z_1} + \frac{1}{z_2} \right)^2
  \Biggr] \; .
  \nonumber
\end{eqnarray}
Here $r_1=1-x_1$, $r_2=1-x_2$, and $z=1-x_1-x_2$.
This expression is consistent with Merenkov \cite{Mer88}, where leading and
next-to-leading logarithms were calculated.

Performing the integration over the photon angles and over the relative photon
energies, taking into account the symmetry factor $1/2!$ for the emitted
photons, one obtains:
\begin{eqnarray}
\label{eq:2-coll-int}
  \lefteqn{
  \frac{e^4}{2!}
  \int \widetilde{\dd k}_1 \, \widetilde{\dd k}_2 \;
  \Theta(x_1 - \eps) \, \Theta(x_2 - \eps) \;
  \delta(x_1 + x_2 - (1-z)) \;
  K_{\mu\nu}^\mathrm{2-coll}
  } \\
  & = &
  \left[
  - \tilde{g}_{\mu\nu} Q_l^2
  + 4 z \left( \tilde{p}_{1\mu} \tilde{p}_{1\nu} \right)
  \right] \times
  \frac{\alpha^2}{8 \pi^2}
  \left[
    P^{\mathrm{(2)}}_\mathrm{log}
    + P^{\mathrm{(2),IR-div.}}_\mathrm{nonlog}
    + P^{\mathrm{(2),IR-fin.}}_\mathrm{nonlog}
  \right] \; .
  \nonumber
\end{eqnarray}
The previously known leading terms containing double and single logarithms
$L_0$ are contained in $P^{\mathrm{(2)}}_\mathrm{log}$.  The nonleading
terms are split into an infrared divergent piece that depends on $\ln
\eps$, and an infrared finite piece.  An outline of the calculation and
expressions are given in appendix~\ref{sec:double-coll-int}.

It is worth to mention that the terms depending on the soft-photon cutoff
parameter $\eps$ in (\ref{eq:2-coll-int}) do factor nicely, as expected
from the usual soft-photon factorization:
\begin{equation}
  \left[
  - \tilde{g}_{\mu\nu} Q_l^2
  + 4 z \left( \tilde{p}_{1\mu} \tilde{p}_{1\nu} \right)
  \right] \times
  \left(\frac{\alpha}{2 \pi}\right)^2 P(z,L_0)
  \cdot 2(L_0-1) \ln \frac{1}{\eps}
  \; .
\end{equation}


\subsubsection{Final state collinear radiation}

Consider now the kinematic region where one photon, say, photon 1, is
emitted almost collinearly to the incoming electron, i.e., $k_1 \simeq x_1
p$, and the other close to the outgoing electron, so that $k_2\simeq
\xi(p_2+k_2)$.  The double Compton tensor then simplifies to
\begin{eqnarray*}
  K_{\mu\nu}^\mathrm{FSR}
  & \simeq &
  \left[
    - \tilde{g}_{\mu\nu} \, \frac{Q_l^2}{1-\xi}
    + 4 (1-x_1) \cdot \left( \tilde{p}_{1\mu} \tilde{p}_{1\nu} \right)
  \right] \times
  \\ && \; {} 2
  \left[
    \frac{1+(1-x_1)^2}{x_1}\frac{1}{z_1} - 2(1-x_1) \frac{m^2}{z_1^2}
  \right] \times
  \\ && \; {} 2
  \left[
    \frac{1+(1-\xi)^2}{\xi}\frac{1}{z_2'} - 2 \frac{m^2}{z_2'{}^2}
  \right]
  \; ,
\end{eqnarray*}
exhibiting the expected complete factorization of collinear initial and
final state radiation, respectively.

Integrating this expression over the emission angles of photon 1, one obtains
\begin{eqnarray*}
  \frac{E^2}{\pi} \int \dd\Omega_1 \;
  K_{\mu\nu}^\mathrm{FSR}
  & = &
  \left[
    - \tilde{g}_{\mu\nu} \, \frac{Q_l^2}{1-\xi}
    + 4 (1-x_1) \cdot \left( \tilde{p}_{1\mu} \tilde{p}_{1\nu} \right)
  \right] \times
  \\ && \; {}
  \frac{2}{x_1} \, P(1-x_1,L_0) \times
  \\ && \; {} 2 \,
  \left[
    \frac{1+(1-\xi)^2}{\xi}\frac{1}{z_2'} - 2 \frac{m^2}{z_2'{}^2}
  \right] \; .
\end{eqnarray*}

Note that we have not yet integrated over the angles of the final state
photon.  The treatment of final state radiation depends on the experimental
situation, i.e., whether the detector is able to resolve a photon collinear
to the electron, or whether it just measures the sum of their energies.


\subsubsection{Semi-collinear emission}
\label{sec:sc}

The final case covers the kinematic range where one photon is emitted
almost collinearly to the incoming electron, while the other photon is
emitted at an angle $\vartheta_2 > \vartheta_0$, but not collinear to the
final electron.  We denote this kinematic domain as the semi-collinear one.

In order to be consistent with the above calculation of the double
collinear emission, we shall perform the angular integration over the
collinear photon and drop all contributions of the order
$\mathcal{O}(\vartheta_0)$ and $\mathcal{O}(\zeta_0^{-1})$.  We find indeed
factorization of initial state radiation for large emission angles of the
second photon, i.e., $\vartheta_2 \gg \vartheta_0$.  However, in the
vicinity of the forward cone that is defined by the solid angle of the PD,
there are contributions from further terms with a complicated
$z_2$-dependence that spoil a naive factorization.  These additional terms
fall off rapidly and essentially contribute only in the small region
$\vartheta_0 < \vartheta_2 \lesssim 2\vartheta_0$.

Assuming that photons in this narrow region outside the PD will not be
measured, we integrate these terms over angles and split their contribution
schematically as follows:
\begin{eqnarray}
  \label{eq:leptonic-semi-coll}
  \lefteqn{
  \frac{E^4}{\pi^2}
  \int\limits_{\vartheta_1<\vartheta_0} \dd\Omega_1
  \int\limits_{\vartheta_2>\vartheta_0} \dd\Omega_2 \;
  K_{\mu\nu}^{\mathrm{semi-coll}}
  }
  \nonumber \\
  & \simeq &
  \frac{E^2}{\pi} \int\limits_{\vartheta_2>\vartheta_0} \dd\Omega_2 \;
  \Biggl\{
  \Biggl[
  - \tilde{g}^{\mu\nu}
    \frac{(r_1(Q_l^2+z_2))^2+(r_1Q_l^2-z_2')^2}{r_1 z_2 z_2'}
  \nonumber
  \\
  && \qquad \qquad \qquad \quad {}
  + 4 r_1^2 \tilde{p}_1^\mu \tilde{p}_1^\nu
    \frac{Q_h^2}{r_1 z_2 z_2'}
  + 4 \tilde{p}_2^\mu \tilde{p}_2^\nu
    \frac{Q_h^2}{r_1 z_2 z_2'}
  \Biggr]
  \cdot
  \frac{4}{x_1 r_1} \, P(r_1,L_0)
  \Biggr\}
  \nonumber \\
  & + &
  \left[
  - \tilde{g}_{\mu\nu} Q_l^2
  + 4 (r_1-x_2) \left( \tilde{p}_{1\mu} \tilde{p}_{1\nu} \right)
  \right] \cdot
  \frac{4}{x_1 x_2} \, H(x_1,x_2)
  \; ,
\end{eqnarray}
with $r_1=1-x_1$.  The explicit expression for the function $H(x_1,x_2)$,
which collects the mentioned non-factorizing, quasi-collinear terms, is
given in appendix~\ref{sec:H}.  It is infrared-finite and does not contain
any logarithm of a large scale.

The integrand in the first part on the r.h.s.\ of
(\ref{eq:leptonic-semi-coll}) can be rewritten as
\begin{equation}
  \Biggl\{ \ldots \Biggr\} \simeq
  \frac{1}{x_1} \, P(r_1,L_0)
  \cdot
  \frac{4}{r_1} B_{\mu\nu}^\mathrm{Born}(r_1 p_1,p_2)
  \; ,
\end{equation}
where for the sake of consistency one should drop terms of order $m^2$ on
the r.h.s.  For a discussion and further details see
ref.~\cite{Anlauf:2002}.

Since in our decomposition of phase space only photon 1 reaches the PD, we
have to identify $r_1$ by $z$ and $x_1$ by $1-z$ in the above expressions.
However, we still need to integrate over the phase space of the other
photon that is emitted at large angles.  This calculation depends on the
complete scattering process and in general requires a numerical
integration.


\section{Radiative DIS}

Let us now turn to the description of the radiative scattering process
\[
  e(p) + p(P) \to
  e(p') + X(P') + \gamma(k_1) \; \left( {} + \gamma(k_2) \right)
  \; .
\]
As argued in the introduction, the leading raditiative corrections to the
process stem from emission of photons off the lepton line.%
\footnote{Another gauge invariant set of large corrections is due to the
running of the QED coupling $\alpha$.  Note that we have already included
it in the expression for the Born cross section, so we won't discuss it
further.}
It is now straightforward to contract the radiatively corrected Compton
tensor of the previous section with the hadron tensor
\begin{eqnarray}
  H_{\mu\nu}(P,q_h)
  &=& 4\pi
  \left(
  - \tilde{g}_{\mu \nu} F_1(x_h,Q^2_h)
  + \tilde{P}_{\mu} \tilde{P}_{\nu} \, \frac{1}{P\cdot q_h} \, F_2(x_h,Q^2_h)
  \right)
  \nonumber \\
  &=& 4\pi
  \left(
  - \tilde{g}_{\mu \nu} F_1(x_h,Q^2_h)
  + \tilde{P}_{\mu} \tilde{P}_{\nu} \, \frac{2x_h}{Q_h^2} \, F_2(x_h,Q^2_h)
  \right)
  \; ,
  \nonumber \\
  \tilde P_{\nu} &=& P_{\nu}-q_{h\nu}\frac{P\cdot q_h}{q_h^2}
  \; , \quad
  x_h=\frac{Q^2_h}{2P\cdot q_h}
  \; , \quad
  Q_h^2 = -q_h^2
  \; .
  \nonumber
\end{eqnarray}
Here $q_h$ denotes the four-momentum transfer to the hadronic system.
Taking into account the emission of two photons, $q_h = p_1 - p_2 - k_1 -
k_2$.

Applying the results from the previous section, we find for the
contribution from virtual and soft corrections to the cross section:
\begin{equation} \label{eq:V+S}
  \frac{1}{\hat{y}} \,
  \frac{\dd^3\sigma_\mathrm{V+S}}{\dd\hat{x}\,\dd\hat{y}\,\dd z} =
  \frac{\alpha^2}{4\pi^2} \left[ P(z,L_0) \tilde{\rho} - T \right]
  \tilde\Sigma(\hat{x},\hat{y},\hat{Q}^2) \; ,
\end{equation}
where $\tilde\rho$ is taken from (\ref{eq:rho-v+s}) with
\begin{eqnarray} \label{eq:Y-c-elastic}
  Y & = & \frac{\Eep}{E_e}
  = z(1 - \hat{y}) + \hat{x}\hat{y} \, \frac{E_p}{E_e}
  \; ,
  \nonumber \\
  c & = & \cos\theta =
  \frac{z(1 - \hat{y}) E_e - \hat{x}\hat{y} E_p}
       {z(1 - \hat{y}) E_e + \hat{x}\hat{y} E_p}
  \; .
\end{eqnarray}

In the calculation of the contributions from the emission of two hard
photons, we decompose the phase space into three regions discussed in the
previous section (see also \cite{AAKM:nlo}): \textit{i)} both hard photons
hit the forward photon detector, i.e., both are emitted within a narrow
cone around the electron beam $(\vartheta_{1,2} \leq \vartheta_0, \,
\vartheta_0 \ll 1)$; \textit{ii)} one photon is tagged in the PD, while the
other is collinear to the outgoing electron $(\vartheta_2' \equiv
\Angle(\vec{k}_2,\vec{p}{\,'}) \leq \vartheta'_0)$; and finally
\textit{iii)} the second photon is emitted at large angles (i.e., outside
the defined narrow cones) with respect to both incoming and outgoing
electron momenta.  For the sake of simplicity, we assume that $m/E_e \ll
\vartheta'_0 \ll 1$.

The contribution from the kinematic region \textit{i)}, with both hard
photons being tagged in the PD, but only the sum of their energies
measured, reads:
\begin{eqnarray} \label{eq:sig-i}
  \frac{1}{\hat{y}} \,
  \frac{\dd^3 \sigma^{\gamma\gamma}_i}{\dd\hat{x}\,\dd\hat{y}\, \dd z}
  & = &
  \frac{\alpha^2}{8\pi^2}
  \left[
    P^{\mathrm{(2)}}_\mathrm{log}
    + P^{\mathrm{(2),IR-div.}}_\mathrm{nonlog}
    + P^{\mathrm{(2),IR-fin.}}_\mathrm{nonlog}
  \right]
  \tilde\Sigma
  \; ,
\end{eqnarray}
see eq.~(\ref{eq:2-coll-int}).

In region \textit{ii)} we need to distinguish between the cases of whether
a photon emitted close to the outgoing electron can be detected separately
(exclusively), or whether its energy and momentum are measured together
with the electron (inclusively), as this affects the reconstructed
kinematic variables.

For the exclusive event selection, when only the scattered electron is
detected, we obtain
\begin{equation} \label{eq:sig-ii-excl}
  \frac{1}{\hat{y}} \,
  \frac{\dd^3 \sigma^{\gamma\gamma}_{ii,\mathrm{excl}}}{\dd\hat{x}\,\dd\hat{y}\, \dd z}
  =
  \frac{\alpha^2}{4\pi^2} P(z,L_0)
  \int\limits_{\zeta_\mathrm{min}}^{\zeta_\mathrm{max}}
  \frac{\dd \zeta}{\zeta^2}
  \left[ \frac{1+\zeta^2}{1-\zeta} \left(\widetilde L-1\right) + 1-\zeta \right]
  \tilde\Sigma_f \; ,
\end{equation}
where $\tilde{L} = \ln (\Eep^2\vartheta_0'{}^2/m^2)=
\ln\left(E_e^2\vartheta_0'{}^2/m^2\right)+2\ln Y$, and
$\tilde\Sigma_f=\tilde\Sigma(x_f,y_f,Q_f^2)$ is an implicit function of
$\zeta$ via the relation between the ``internal'' kinematic variables
$x_f,y_f,Q_f^2$ and the ``external'' ones $\hat{x}, \hat{y}, \hat{Q}^2$
(see refs.~\cite{AAKM:JETP,AAKM:nlo,Anl99:Sigma} for more details).  The
integration limits explicitly depend on the method for the determination of
kinematic variables.

In the case of a calorimetric event selection, where only the sum of the
energies of the outgoing electron and collinear photon is measured and
taken into account in the determination of the kinematic variables, the
corresponding contribution reads
\begin{equation} \label{eq:sig-ii-cal}
  \frac{1}{\hat{y}} \,
  \frac{\dd^3 \sigma^{\gamma\gamma}_{ii,\mathrm{cal}}}{\dd\hat{x}\,\dd\hat{y}\, \dd z}
  =
  \frac{\alpha^2}{4\pi^2} P(z,L_0)
  \int\limits_0^{\zeta_\mathrm{max}}
  \!\! \dd \zeta
  \left[ \frac{1+\zeta^2}{1-\zeta}
    \left(\tilde{L} - 1 + 2\ln\zeta \right) + 1-\zeta \right]
  \tilde\Sigma \; ,
\end{equation}
see also \cite{AAKM:JETP,AAKM:nlo,Anl99:Sigma}.

For the contribution from the semi-collinear region \textit{iii)} we apply
the results from section \ref{sec:sc} to obtain:
\begin{eqnarray} \label{eq:sig-iii}
  \frac{1}{\hat{y}} \,
  \frac{\dd^3 \sigma^{\gamma\gamma}_{iii}}{\dd\hat{x}\,\dd\hat{y}\, \dd z}
  & = &
  \frac{\alpha^2}{\pi^2} P(z,L_0)
  \int\frac{\dd^3 k_2}{|\vec{k}_2|}
  \, \frac{\alpha^2(Q_h^2)}{Q_h^4}
  \, I^{\gamma}(zp,p',k_2)
  \nonumber \\
  & + &
  \frac{\alpha^2}{4\pi^2}
  \int\limits_{\eps}^{x_2^t}
  \dd x_2 \,
  \frac{z}{z-x_2} \, H(1-z,x_2) \, \tilde\Sigma(x_t,y_t,Q_t^2)
  \; ,
\end{eqnarray}
where the `radiation kernel' $I^{\gamma}$ and the boundaries of integration
for the logarithmic part are given in \cite{AAKM:nlo}.  For the second,
quasi-collinear contribution, we have used the abbreviations
\begin{equation}
  x_t = \frac{(z-x_2) \hat{x}\hat{y}}{z\hat{y}-x_2}
  \; , \quad
  y_t = \frac{z\hat{y}-x_2}{z-x_2}
  \; , \quad
  Q_t^2 = \hat{Q}^2 \frac{z-x_2}{z}
  \; ,
\end{equation}
and the upper limit of integration is
\begin{equation}
  \label{eq:x2t}
  x_2^t = z \hat{y} \; .
\end{equation}

The total contribution from QED radiative corrections is finally obtained
by adding up (\ref{eq:V+S}), (\ref{eq:sig-i}), (\ref{eq:sig-iii}), and,
depending on the chosen event selection, (\ref{eq:sig-ii-excl}) or
(\ref{eq:sig-ii-cal}).  One easily verifies that the unphysical IR
regularization parameter $\eps$ cancels in the sum.


\section{Results and Discussion}
\label{sec:results}

In this section we shall present numerical results obtained from the above
expressions and compare to next-to-leading radiative corrections
\cite{AAKM:nlo}.  As input we used
\begin{equation}
  E_e = 27.5 \GeV \, , \quad
  E_p = 820  \GeV \, , \quad
  \vartheta_0 = 0.5 \; \mathrm{mrad} \, .
\end{equation}
We chose the ALLM97 parameterization \cite{ALLM97} as structure function
with $R=0$, no cuts were applied to the phase space of the second photon,
and we assumed a calorimetric event selection.  Furthermore, we took a
fixed representative angular resolution of
$\vartheta_0'=50\,\mathrm{mrad}$.


\begin{figure}
  \begin{center}
    \begin{picture}(120,120)
      \put(0,0){
        \includegraphics[width=120mm]{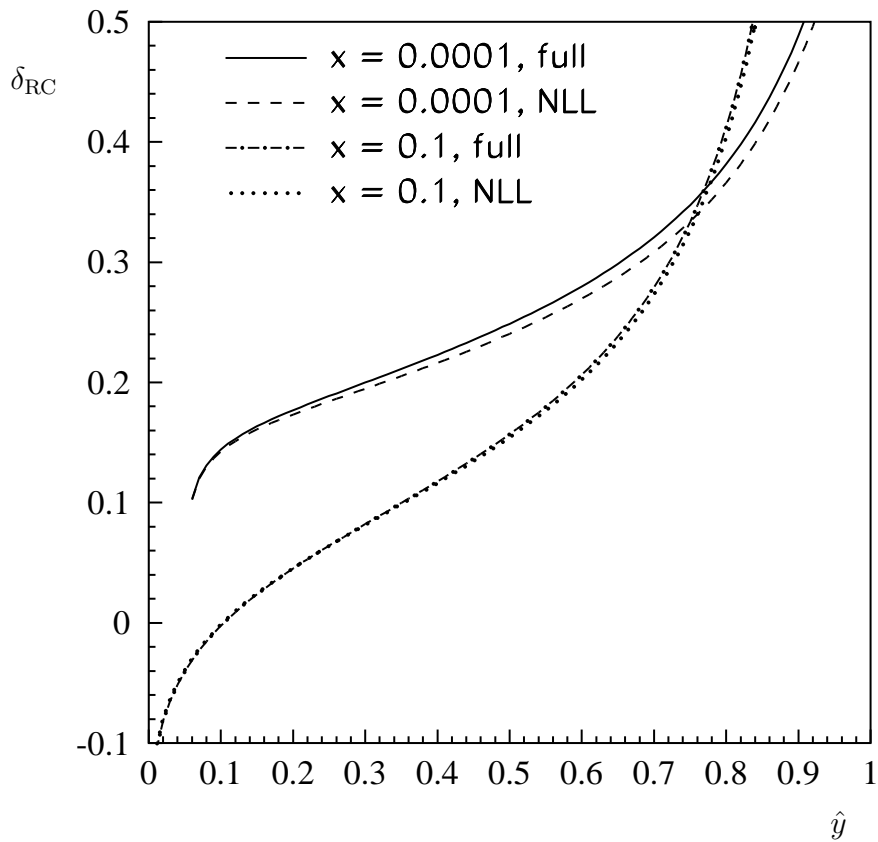}
        }
      \put(104,0){$\hat{y}$}
      \put(-5,99){$\delta_\mathrm{RC}$}
    \end{picture}
    \caption{Radiative corrections $\delta_\mathrm{RC}$
      (\protect\ref{eq:delta}) with full leptonic $\mathcal{O}(\alpha)$
      vs.\ next-to-leading logarithmic accuracy at $\hat{x} = 10^{-4}$ and
      $\hat{x} = 0.1$ and a tagged photon energy of 20\GeV.  No cuts have
      been applied to the phase space of the second (semi-collinear)
      photon.}
    \label{fig:plot1}
  \end{center}
\end{figure}


Figure~\ref{fig:plot1} compares the radiative correction
\begin{equation} \label{eq:delta}
  \delta_\mathrm{RC} =
  \frac{\dd^3\sigma}{\dd^3\sigma_\mathrm{Born}} - 1 \; ,
\end{equation}
calculated with next-to-leading logarithmic and full $\mathcal{O}(\alpha)$
accuracy for the electron method at $\hat{x}=10^{-4}$ and $\hat{x}=0.1$ and
for a tagged energy of $E_\mathrm{PD} = 20\GeV$.  Similar to the well-known
QED corrections to DIS (see e.g., \cite{BBC+97} and references cited
therein), the corrections are large and positive for large $\hat{y}$, while
they may become large and negative for $\hat{y} \to 0$ at large $\hat{x}$.

Furthermore, one sees that the difference between the full result and the
one at next-to-leading logarithmic accuracy \cite{AAKM:nlo} is very small;
it is typically at the permille level except for large $\hat{y}$, where the
phase space for the `lost photon' gets large, see (\ref{eq:x2t}), and for
small $\hat{x}$.  In this case it easily reaches values of the order of a
percent.  This may be important in view of the increased statistics
expected at the upgraded HERA collider.

Looking at the individual contributions beyond the next-to-leading
logarithmic approximation, one finds a significant cancellation between
terms from virtual+soft corrections and from double collinear emission.
Their sum is typically of the order permille, and it depends only on $z$,
c.f.\ (\ref{eq:V+S}) and (\ref{eq:sig-i}).  Therefore the total difference
can be mainly attributed to the contribution of photons emitted into a
small region outside but close to the PD.

To summarize, we have calculated the full leptonic QED corrections to DIS
with tagged initial state radiation.  We find that the corrections beyond
next-to-leading logarithmic accuracy are typically very small, but can
still be significant, i.e., of the order of a percent, in the interesting
region of small $x$, where the use of radiative events appears most
promising.




\appendix


\section{Integrals for double collinear emission}
\label{sec:double-coll-int}

The calculation of the contribution from double collinear emission assumes
that only the sum of the photon energies, $(1-z)E$, will be measured in the
forward photon detector.

We split the integration of expression (\ref{eq:2-coll-int}) over the
restricted two-photon phase space in the following way:
\begin{eqnarray}
  \label{eq:2-coll-ps}
  \lefteqn{
  \int \widetilde{\dd k}_1 \; \widetilde{\dd k}_2 \;
  \Theta(\vartheta_0 - \vartheta_1) \,
  \Theta(\vartheta_0 - \vartheta_2) \;
  \delta\left( (1-z) - (x_1+x_2) \right)
  \Bigl[ \ldots \Bigr]
  } \nonumber \\
  & = &
  \frac{1}{(4\pi)^4}
  \int x_1 \, \dd x_1 \int x_2 \, \dd x_2 \;
  \delta\left( (1-z) - (x_1+x_2) \right)
  \overline{\Bigl[ \ldots \Bigr]}
  \; .
\end{eqnarray}
In the last line we adopted the notation of Arbuzov et
al.~\cite{Arbuzov:1997qb},
\begin{eqnarray}
  \overline{\Bigl[ \ldots \Bigr]}
  & := &
  \frac{E^4}{\pi^2}
  \int \dd\Omega_1 \, \dd\Omega_2 \;
  \Theta(\vartheta_0 - \vartheta_1) \,
  \Theta(\vartheta_0 - \vartheta_2) \;
  \Bigl[ \ldots \Bigr]
  \; ,
\end{eqnarray}
for the angular part of the integrals.


\subsection{Integrals over photon angles}

We shall now provide the relevant angular integrals needed for the
contribution of two photons emitted almost collinearly to the incoming
electron.
The calculation is performed under the assumption that
$E^2\vartheta_0^2/m^2 \gg 1$ and $\vartheta_0^2 \ll 1$.  For details we
refer the reader to \cite{Anlauf:2002}.

Using the abbreviation
\[
  L_0 = \ln \frac{E^2\vartheta_0^2}{m^2} \; ,
\]
we obtain:
\begin{eqnarray}
  \overline{ \left[ \frac{1}{z_1 z_2} \right] }
  & = &
  \frac{1}{x_1 x_2} \, L_0^2 \; ,
  \nonumber \\
  \overline{ \left[ \frac{m^2}{z_1^2 \Delta} \right] }
  & = &
  \frac{1}{x_1^2 x_2 r_1} \left[
    L_0 +
    \ln \frac{ x_2 r_1 }{1-z}
  \right]
  -
  \frac{z}{x_1 x_2^2 r_1}
  \ln \frac{r_1 (1-z)}{ x_1 z } \; ,
  \nonumber
  \\
  \overline{ \left[ \frac{1}{z_1 \Delta} \right] }
  & = &
  \frac{1}{x_1 x_2 r_1}
  \Biggl[
  \frac{1}{2} L_0^2
  + L_0 \ln \frac{ x_2 r_1^2}{x_1 z}
  + \dilog \left( - \frac{x_2}{x_1 z} \right)
  + \Xi\left( \cos\psi; \frac{x_1 r_2}{x_2 r_1} \right)
  \Biggr]
  \; ,
  \nonumber
  \\
  \overline{ \left[ \frac{z_2}{z_1 \Delta^2} \right] }
  & = &
  \frac{1}{x_1 x_2 r_1^2}
  \Biggl\{
  \frac{1}{2} L_0^2
  + L_0
    \left[ \ln \frac{ x_2 r_1^2}{x_1 z} - 1 + \frac{x_1x_2}{z} \right]
  \nonumber
  \\ && \qquad \qquad {}
   + \dilog \left( - \frac{x_2}{x_1 z} \right)
   + \Xi\left( \cos\psi; \frac{x_1 r_2}{x_2 r_1} \right)
  \\ && \qquad \qquad {}
  + \frac{r_1 r_2-2z}{z} \ln (2x_2)
  + \frac{r_1(r_2-2z)}{z} \ln \frac{x_1}{1-z}
  \nonumber
  \\ && \qquad \qquad {}
  + \frac{r_1(3r_2-4z)}{2z} \ln z
  - \frac{r_1r_2+4x_1z}{2z} \ln r_1
  - \frac{r_1 r_2}{2z}   \ln r_2
  \nonumber
  \\ && \qquad \qquad {}
  + \frac{4z-r_1r_2}{2z}
    \ln \left( \eta + x_2 r_1 + x_1 r_2 \cos\psi \right)
  \nonumber
  \\ && \qquad \qquad {}
  - \frac{r_1r_2}{2z}
    \ln \left( \eta + x_1 r_2 + x_2 r_1 \cos\psi \right)
  \Biggr\}
  \; ,
  \nonumber
\end{eqnarray}
where $z=1-x_1-x_2$, $r_{1,2}=1-x_{1,2}$, and
\begin{eqnarray}
  \label{def:Xi}
  \cos\psi & = &
  1 - \frac{2 x_1 x_2}{r_1 r_2}
  \; ,
  \nonumber \\
  \eta & = &
  \sqrt{(x_1+x_2)(x_1+x_2-4x_1x_2)}
  \; ,
  \nonumber
  \\
  \Xi(t;x)
  & = &
  \frac{1}{2} \ln^2 \left( \frac{ \sqrt{1+2tx+x^2} + tx + 1 }{2} \right)
  \\ &+&
  \dilog\left( \frac{(1+t)x}{ \sqrt{1+2tx+x^2} + tx + 1 } \right)
  \nonumber
  \\ &+&
  \dilog\left( - \frac{(1-t)x}{ \sqrt{1+2tx+x^2} + tx + 1 } \right)
  \; .
  \nonumber
\end{eqnarray}
The integrals that contribute only non-enhanced terms (i.e., no $L_0$'s)
are:
\begin{eqnarray}
  \overline{ \left[ \frac{m^2}{z_1 z_2 \Delta} \right] }
  & = &
  \frac{1}{x_1^2 x_2^2}
  \Biggl[
   (1-z) \ln (1-z) + z \ln z
   - r_1 \ln r_1 - r_2 \ln r_2
  \nonumber \\ && \qquad \quad {}
   - x_1 \ln x_1 - x_2 \ln x_2
  \Biggr]
  \; ,
  \nonumber \\
  \overline{ \left[ \frac{m^2}{z_1 \Delta^2} \right] }
  & = &
  \frac{1}{x_1 x_2^2} \ln \frac{r_1(1-z)}{x_1 z}
  \; ,
  \nonumber \\
  \overline{ \left[ \frac{m^4}{z_1^2 \Delta^2} \right] }
  & = &
  \frac{1}{x_1^2 x_2^2}
  \left[
    1 - \frac{x_1 y}{x_2} \ln \frac{r_1(1-y)}{x_1 y}
  \right]
  \; ,
  \\
  \overline{ \left[ \frac{m^4}{z_1 z_2 \Delta^2} \right] }
  & = &
  \frac{1}{6 x_1^3 x_2^3}
  \Biggl[
      x_1^2 (3-2x_1) \ln \frac{(1-z)r_1}{x_1 z}
    + x_2^2 (3-2x_2) \ln \frac{(1-z)r_2}{x_2 z}
  \nonumber \\ && \qquad \quad {}
    + \ln \frac{z}{r_1 r_2} - 2 x_1 x_2
  \Biggr]
  \; .
  \nonumber
\end{eqnarray}
The remaining integrals can be obtained from those given above by
exchanging photons 1 and 2, i.e., $x_1 \leftrightarrow x_2$ and $r_1
\leftrightarrow r_2$.

It should be noted that the coefficients of the double and single logarithmic
terms ($L_0^2$ and $L_0$) agree with refs.~\cite{Mer88,Arbuzov:1997qb}.


\subsection{Integrals over relative photon energy}

Having performed the angular integration, we still need to integrate over
the relativ photon energy, see (\ref{eq:2-coll-ps}).  As this integral will
be infrared-divergent, we introduce a soft-photon cutoff $\eps$ on the
minimum energy fraction of each photon, which is identical to the one used
in the soft-photon contribution.  We thus define:
\begin{eqnarray}
  \left\langle
  \Bigl[
  \cdots
  \Bigr]
  \right\rangle
  & := &
  \int\limits_\eps^1 x_1 \, \dd x_1 \int\limits_\eps^1 x_2 \, \dd x_2 \;
  \delta\left( (1-z) - (x_1+x_2) \right)
  \Bigl[
  \cdots
  \Bigr]
\end{eqnarray}

With the substitution $ x_1 \to (1-z) \cdot u $ and the abbreviation
$\tilde\eps=\eps/(1-z)$, we have, after elimination of the trivial
$\delta$-function:
\begin{eqnarray*}
  \left\langle
  \Bigl[
  \cdots
  \Bigr]
  \right\rangle
  & = &
  (1-z)
  \int\limits_{\tilde\eps}^{1-\tilde\eps} \dd u \;
  \, x_1 \, x_2
  \Bigl[
  \cdots
  \Bigr]_{x_1 = (1-z) u, x_2 = (1-z)(1-u), \ldots}
\end{eqnarray*}

For the logarithmically ($L_0$) enhanced leading terms, we find the familar
result:
\begin{eqnarray*}
  P^{\mathrm{(2)}}_\mathrm{log}
  & = &
  \left[
    - 4 \frac{1+z^2}{1-z} \ln \tilde{\eps}
    + (1+z) \ln z - 2(1-z)
  \right] L_0^2
  \\
  &+&
  \left[
    6(1-z)
    + \frac{3+z^2}{1-z} \ln^2 z
    + 4 \frac{(1+z)^2}{1-z} \ln \tilde{\eps}
  \right] L_0
  \\
  & = &
  \left[
    P_\Theta^{(2)}(z)
    + 2 \, \frac{1+z^2}{1-z}
    \left( \ln z - \frac{3}{2} - 2 \ln \eps \right)
  \right] L_0^2
  \\
  &+&
  \left[
    6(1-z)
    + \frac{3+z^2}{1-z} \ln^2 z
    + 4 \frac{(1+z)^2}{1-z} \ln \frac{\eps}{1-z}
  \right] L_0
  \; ,
\end{eqnarray*}
with the leading-log radiator
\begin{eqnarray*}
  P_\Theta^{(2)}(z)
  & = &
  2 \left[
    \frac{1+z^2}{1-z} \left( 2 \ln(1-z) - \ln z + \frac{3}{2} \right)
    + \frac{1}{2} (1+z) \ln z - 1 + z
  \right] \; .
\end{eqnarray*}

The analytical calculation of the remaining, non-enhanced terms is quite
tedious, leading to lengthy expressions involving many dilogarithms and
trilogarithms (see e.g., \cite{Lewin:1981}).

As these terms also contain IR-divergent contributions, we shall pursue
here the following approach.  We analytically extract those terms in the
integrand that either contribute to the infrared-divergence as $\eps \to 0$
or survive in the limit $z \to 1$, before performing the integral over the
remaining expression numerically.  Besides, this separation improves the
stability of the numerical integration.

For the IR-divergent pieces of the non-enhanced terms we find
\begin{eqnarray}
  P_\mathrm{nonlog}^\mathrm{IR-div}
  & = &
  \left\langle
    \frac{4z}{(x_1 x_2)^2}
  \right\rangle
  =
  \frac{4z}{1-z}
  \int\limits_{\tilde\eps}^{1-\tilde\eps} \dd u \;
  \frac{1}{u(1-u)}
  \simeq \frac{8z}{1-z} \ln \tilde\eps
  \; .
\end{eqnarray}
We then decompose the infrared-finite pieces as follows:
\begin{equation}
  P^{\mathrm{(2),IR-fin.}}_\mathrm{nonlog}(z)
  = P_\mathrm{nonlog}^{z\to 1} + P_\mathrm{nonlog}^\mathrm{rem}(z) \; ,
\end{equation}
so that $P_\mathrm{nonlog}^\mathrm{rem}(z=1)=0$.  The first term on the
r.h.s., obtained in the limit $z \to 1$ reads:
\begin{eqnarray}
  P_\mathrm{nonlog}^{z\to 1}
  & = &
  \Biggl\langle
    \frac{8}{3 x_1 x_2}
    \Biggl[
      \frac{x_1 + 3x_2}{x_2^2} \ln x_1
    + \frac{3x_1 + x_2}{x_1^2} \ln x_2
  \nonumber \\ && \qquad \qquad {}
    - \frac{(1-z)^3}{x_1^2 x_2^2} \ln (1-z)
    + \frac{1-z}{x_1 x_2}
    \Biggr]
  \Biggr\rangle
  \nonumber \\
  & = &
  \int\limits_{\tilde\eps}^{1-\tilde\eps} \dd u \;
  \frac{8}{3}
  \left[
    \frac{1}{u(1-u)}
    + \frac{3-2u}{(1-u)^2} \ln u
    + \frac{1+2u}{u^2} \ln (1-u)
  \right]
  \nonumber \\
  & = & - \frac{16}{9} \left( 3 + \pi^2 \right)
  \; ,
\end{eqnarray}
where we neglected terms of order $\eps$.

\begin{figure}
  \begin{center}
    \begin{picture}(120,70)
      \put(0,0){\includegraphics{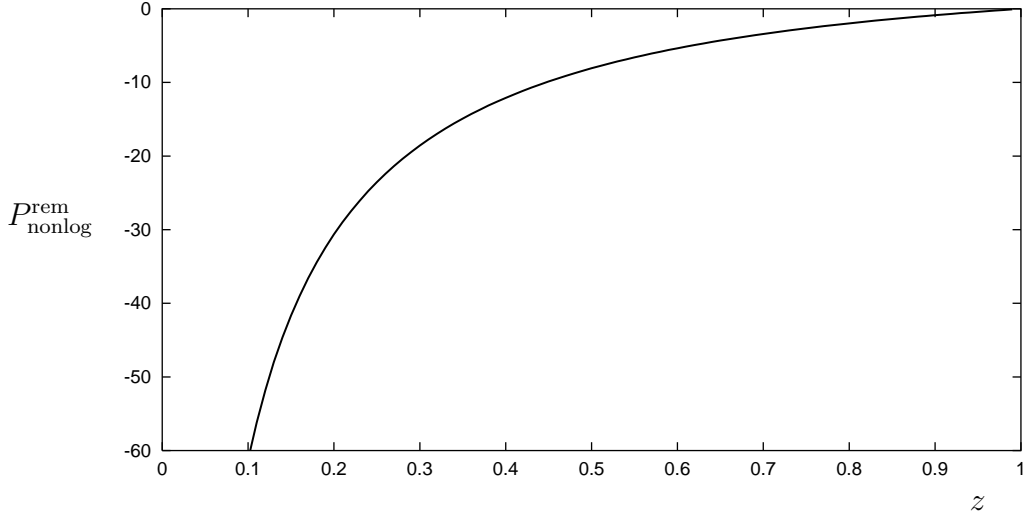}}
      \put(116,-3){$z$}
      \put(-12,35){$P_\mathrm{nonlog}^\mathrm{rem}$}
    \end{picture}
    \caption{The infrared-finite, `normalized' part of the double-collinear
      contribution, $P_\mathrm{nonlog}^\mathrm{rem}(z)$.}
    \label{fig:Pnonlog}
  \end{center}
\end{figure}

Finally, we plot in figure~\ref{fig:Pnonlog} the `remainder'
$P_\mathrm{nonlog}^\mathrm{rem}(z)$, which we evaluated numerically.


\section{Non-factorizing quasi-collinear piece}
\label{sec:H}

In the discussion of the contributions to the semi-collinear piece, a
non-factorizing part was split off in eq.~(\ref{eq:leptonic-semi-coll}).
The function $H$ appearing there is given by
\begin{eqnarray}
  H(x_1,x_2)
  & = &
  \frac{r_1^3+\chi r_2}{x_1 x_2 r_1} \, H_1
  + \frac{r_2^3+\chi r_1}{x_1 x_2 r_2} \, H_2
  - \chi \left(
  \frac{H_3}{r_1^2} + \frac{H_4}{r_2^2}
  \right)
  \; ,
\end{eqnarray}
with
\begin{eqnarray*}
  H_1 & = &
  - \Xi\left(\cos\psi, \frac{x_1 r_2}{x_2 r_1} \right)
  \; ,
  \\
  H_2 & = &
  \frac{1}{2} \ln^2 \frac{x_1 r_2^2}{x_2 \chi}
  + \frac{\pi^2}{6}
  + \dilog\left( - \frac{x_1 x_2}{\chi} \right)
  - \Xi \left(\cos\psi, \frac{x_2r_1}{x_1r_2} \right)
  \; ,
  \\
  H_3 & = &
  H_1 +
  \frac{2\cos\psi}{1+\cos\psi}
  \ln (2 x_2 r_1)
  - \frac{1}{1+\cos\psi}
  \ln \frac{1+\cos\psi}{2}
  \\
  & - &
  \frac{1+2\cos\psi}{1+\cos\psi} \ln ( \eta + x_2 r_1 + x_1 r_2 \cos\psi)
  \\
  & + &
  \frac{1}{1+\cos\psi} \ln ( \eta + x_1 r_2 + x_2 r_1 \cos\psi)
  \; ,
  \\
  H_4 & = &
  H_2 +
  \frac{2\cos\psi}{1+\cos\psi} \ln [(1+\cos\psi) x_2 r_1 ]
  + \frac{1}{1+\cos\psi}       \ln \frac{1+\cos\psi}{2}
  \\
  & + &
  \frac{1}{1+\cos\psi} \ln ( \eta + x_2 r_1 + x_1 r_2 \cos\psi)
  \\
  & - &
  \frac{1+2\cos\psi}{1+\cos\psi} \ln ( \eta + x_1 r_2 + x_2 r_1 \cos\psi)
  \; ,
  \\
  \chi & = & 1-x_1-x_2
  \; , \quad
  r_{1,2} = 1-x_{1,2}
  \; .
\end{eqnarray*}
The function $\Xi$ and expressions $\eta$ and $\cos\psi$ are defined in
(\ref{def:Xi}).  For a derivation and details we refer the reader to
\cite{Anlauf:2002}.




\begin{thebibliography}{88}

\bibitem{DIS99:Arkadov}
V.~Arkadov [H1 Collaboration],
Nucl.\ Phys.\ Proc.\ Suppl.\  {\bf 79} (1999) 179.

\bibitem{Adloff:2001qk}
C.~Adloff {\it et al.}  [H1 Collaboration],
Eur.\ Phys.\ J.\ C {\bf 21} (2001) 33
[hep-ex/0012053].

\bibitem{DIS01:Dubak}
	A. Dubak [H1-Collaboration], presentation at the
	DIS 2001 International Workshop on Deep Inelastic Scattering,
	Bologna, 27 April - 1 May, 2001,
	and H1-prelim-01-053.

\bibitem{EPS01:ZEUS}
	ZEUS Collaboration,
	\textit{The ZEUS NLOQCD fit to determine parton distribution
	  functions and $\alpha_s$},
	contributed paper no.~628 to the EPS 2001 conference
	(Budapest, Hungary, July 2001).

\bibitem{Gehrmann:2000vu}
T.~Gehrmann,
Phys.\ Lett.\ B {\bf 480} (2000) 77
[hep-ph/0003156].

\bibitem{KPS92}
M.~W.~Krasny, W.~P\l{}aczek and H.~Spiesberger,
Z.\ Phys.\ C {\bf 53} (1992) 687.

\bibitem{Klein:1998mz}
        M. Klein [H1 collaboration], talk no.~404 given at the
        ICHEP98 conference, (Vancouver, Canada, July 1998), and contributed
        paper no.~535

\bibitem{Issever:thesis}
	\c{C}. \.I\c{s}sever,
	\textit{Messung der Protonstrukturfunktionen $F_2(x,Q^2)$ und
	$F_L(x,Q^2)$ bei HERA in radiativer $ep$ Streuung},
	Ph.D. thesis, Dortmund, December 2000;
	DESY-THESIS-2001-032.

\bibitem{Issever:DIS01}
	\c{C}. \.I\c{s}sever [H1 collaboration], presentation at the
	DIS 2001 International Workshop on Deep Inelastic Scattering,
	Bologna, 27 April - 1 May, 2001,
	and H1-prelim-01-042.

\bibitem{H1:2001rad}
	H1 Collaboration,
	\textit{Measurement of the Proton Structure Function using
	Radiative Events at HERA},
	contributed paper to the EPS 2001 (Budapest, Hungary, July 2001)
	and LP 2001 (Rome, Italy, July 2001) conferences.

\bibitem{H1:rad}
T.~Ahmed {\it et al.}  [H1 Collaboration],
Z.\ Phys.\ C {\bf 66} (1995) 529.

\bibitem{ZEUS96}
M.~Derrick {\it et al.}  [ZEUS Collaboration],
Z.\ Phys.\ C {\bf 69} (1996) 607
[hep-ex/9510009].

\bibitem{FGMZ96}
L.~Favart, M.~Gruw\'e, P.~Marage and Z.~Zhang,
Z.\ Phys.\ C {\bf 72} (1996) 425
[hep-ph/9606465].

\bibitem{BKR97}
D.~Y.~Bardin, L.~Kalinovskaya and T.~Riemann,
Z.\ Phys.\ C {\bf 76} (1997) 487
[hep-ph/9612203].

\bibitem{AAKM:ll}
H.~Anlauf, A.~B.~Arbuzov, E.~A.~Kuraev and N.~P.~Merenkov,
Phys.\ Rev.\ D {\bf 59} (1999) 014003
[hep-ph/9711333].

\bibitem{AAKM:JETP}
H.~Anlauf, A.~B.~Arbuzov, E.~A.~Kuraev, N.~P.~Merenkov,
JETP Lett.\  {\bf 66} (1997) 391
[Erratum-ibid.\  {\bf 67} (1997) 305].

\bibitem{AAKM:nlo}
H.~Anlauf, A.~B.~Arbuzov, E.~A.~Kuraev and N.~P.~Merenkov,
JHEP {\bf 9810} (1998) 013
[hep-ph/9805384].

\bibitem{Anl99:Sigma}
H.~Anlauf,
Eur.\ Phys.\ J.\ C {\bf 9} (1999) 69
[hep-ph/9901258].

\bibitem{Spiesberger:1995dm}
H.~Spiesberger,
Phys.\ Rev.\ D {\bf 52} (1995) 4936
[hep-ph/9412286].

\bibitem{KMF87}
E.~A.~Kuraev, N.~P.~Merenkov and V.~S.~Fadin,
Yad.\ Fiz.\  {\bf 45} (1987) 782
[Sov.\ J.\ Nucl.\ Phys.\  {\bf 45} (1987) 486].

\bibitem{Anlauf:2002}
	H. Anlauf, in preparation.

\bibitem{Mer88}
N.~P.~Merenkov,
Sov.\ J.\ Nucl.\ Phys.\  {\bf 48} (1988) 1073
[Yad.\ Fiz.\  {\bf 48} (1988) 1782].

\bibitem{ALLM97}
        H. Abramowicz, A. Levy, DESY 97-251, 1997,
	arXiv:hep-ph/9712415.

\bibitem{BBC+97}
D. Bardin, J. Bl\"{u}mlein, P. Christova, L. Kalinovskaya and T. Riemann,
Acta Phys.\ Polon.\ B {\bf 28} (1997) 511
[arXiv:hep-ph/9611426].

\bibitem{Arbuzov:1997qb}
A.~B.~Arbuzov, V.~A.~Astakhov, E.~A.~Kuraev, N.~P.~Merenkov, L.~Trentadue and E.~V.~Zemlyanaya,
Nucl.\ Phys.\ B {\bf 483} (1997) 83
[hep-ph/9610228].

\bibitem{Lewin:1981}
	L. Lewin,
	\textit{Polylogarithms and Associated Functions},
	North Holland, 1981.

\end{thebibliography}
\end{document}